# The order parameter of high temperature superconductors, as measured by Andreev Saint-James reflection, Raman and neutron scattering

S. Hüfner [1), 2), a)], F. Müller [1)]

[1)] Department of Experimental Physics, Saarland University, Campus C6.3, 66123 Saarbrücken, Germany
[2)] Department of Physics and Astronomy, University of British Columbia, Vancouver, British Columbia, V6T 1Z1 Canada

[a)] Corresponding Author: huefner@mx.uni-saarland.de

By comparing experiments that use Andreev Saint-James reflection, Raman scattering ($B_{2g}$) and neutron scattering on the (cuprate) high temperature superconductors, evidence is found, that all of the techniques can measure the order parameter of these superconductors. The order parameter is the energy needed to take a superconducting boson out of the condensate. It has the same temperature and doping dependence as the superconducting transition temperature $T_c$. This condensation energy is different from the pairing energy (as measured by the pseudogap) except for the doping $x = 0.27$, where the two energies coincide: this latter situation represents the BCS case, which in the view presented here is a special case in the BCS-BEC cross-over situation appropriate for the (cuprate) high temperature superconductors.

**PACS**   74.20.-z (Theories and models of superconducting state)

## 1 Introduction

The (Cuprate) High Temperature Superconductors (CHTSC) still are an interesting subject of research even 20 years after their discovery by Bednorz and Müller [1]. Some of the problems unsolved so far are:
Q(i): What is the coupling mechanism in the superconducting bosons?
Q(ii): Do the CHTSC represent conventional BCS superconductors or are they in the cross-over regime to a Bose Einstein condensate (BEC)?
Q(iii): How are the pseudogap and the superconducting gap related?
Q(iv): Does the pseudogap represent a precursor state to superconductivity or is it the manifestation of competing order?
Q(v): What is the order parameter of the CHTSC, and what are its properties?
In the present communication the points Q(iii) and Q(v) will be dealt with. Finding answers to the questions listed above may help to support a view of the CHTSC that has been developed in Refs. [2-22], although there are also other approaches [23] for their description.

In the picture used here to describe the CHTSC, the charge carriers are the holes introduced by doping and they form two particle singlet states [16-21]. These states are represented by the pseudogap, which has an approximately linear doping dependence. The energy of these bosons multiplied with the charge carrier density (which rises approximately linearly with doping) gives the superconducting order parameter (gap) [5, 7, 8, 22]. The product of the two linear quantities as a function of doping results in the parabolic dependence of the order parameter on the doping as observed in experiment [24]. The bosons formed at the pseudogap temperature $T^*$ condense at $T_c$, thus putting the CHTSC in the BCS-BEC cross-over regime [15].
This picture of the CHTSC is by no means generally accepted [23]. In order to validate it, existing experimental data have to be evaluated against it.
In this communication it is tried to shed more light on the question of the two energy scales observed in the CHTSC, namely the pseudogap (pg) and the superconducting gap (sc), with respect to experiments using Andreev Saint-James (ASJ) reflection, Raman scattering (RS) and inelastic neutron scattering (INS) data. The suggestion will be put forward that the $B_{2g}$ RS resonance [25-28] and the INS magnetic resonance [29] measure the order parameter, as also obtained by ASJ experiments [3, 4]. This provides support to the two gap picture of the CHTSC.

## 2 Outline of the problem

The standard method to measure a gap in a conventional BCS superconductor is by tunnelling or photoemission spectroscopy [30-35]. In one-electron spectroscopies, such as photoemission spectroscopy (ARPES) [35], Normal-Insulator-Superconductor (NIS) and Superconductor-Insulator Superconductor (SIS) junction tunnelling experiments [30, 32-34] as well as STM spectroscopy [31], a pair of electrons is broken up and the emerging electron probes the energy of the electron left behind. In this way, the gap energy $\Delta_{sc}$ (or $2\Delta_{sc}$ in SIS) is measured.

In a BCS superconductor, the gap measured by a one electron spectroscopy like ARPES and measured by a two electron spectroscopy like ASJ is the same, because the coupling energy and the condensation energy are the same (the terms one and two electron spectroscopy will be detailed later).

However, for the CHTSC, two excitations have been observed. Taking a case in point, namely systems with $T_c$ = 95 K at optimal doping, one of these excitations at $2\Delta_{pg}$ = 70 ± 10 meV has a linear doping dependence and is independent of temperature up to $T^* > T_c$. This excitation has been experimentally observed in NIS tunnelling [33], SIS tunnelling [32], STM [31], ARPES [35] and $B_{1g}$ RS experiments [25-28] as a kink in the nodal ARPES dispersion [36, 37], in thermodynamic experiments and in NMR [38]. The gap, as measured by these techniques, starts in lowering the temperature to develop at a temperature $T^*$ above $T_c$ [38] and the magnitude of the gap does not change appreciably in going through the superconducting transition temperature. This gap is called the pseudogap and represents a pairing gap [38, 39]. The same gap, as far as energy is concerned, observed below $T_c$ [31] is often called the superconducting gap, and sometimes the notion is used that the pseudogap changes smoothly into the superconducting gap in going through the superconducting transition temperature [13, 31]. This is not easy to understand, because this interpretation of the experimental data implies a different doping dependence for the transition temperature (parabolic) and the superconducting gap (linear).

A second excitation observed in the CHTSC (40 ± 5 meV for $T_c$ = 95 K systems, OP) is seen in ASJ [3, 4], INS [29], optical absorption [40, 41], $B_{2g}$ RS [25, 26, 28], as a dip in the ARPES spectra [35, 42], as a dip in the SIN and SIS tunnelling spectra [32-34] and as a kink in the antinodal ARPES dispersion [35, 43, 44]. The interpretation of these excitations is controversial, and it is the intention of this communication to shed some light on this problem.

In all the very different experiments the excitations at ~70 meV and ~40 meV have distinct but different temperature and doping dependencies. Therefore, the underlying assumption of the present study is, that there are two gaps in the CHTSC (70 meV vs. 40 meV) that are probed by different techniques. It is suggested that one-electron spectroscopies on CHTSC´s, such as ARPES or STM tunnelling, probe the pseudogap (above and below $T_c$) by breaking-up a pair of electrons and by measuring the energy of the electron left behind. On the other hand, the superconducting energy (order parameter) at 40 meV, which is the energy needed to transfer a boson out of the condensate, is a two-electron transition, which cannot be observed directly by a one-electron technique.

In the picture advocated here, the pseudogap pairs are the preformed pairs that condense below $T_c$ into the superconducting state. Therefore in principle the techniques that determine the superconducting energy, not relying on the superconducting nature of the pairs, should also be able to measure the pseudogap energy. This means that INS should also be able to measure the pseudogap energy however ASJ not. This statement with respect to ASJ is not qwite correct and this point will be discussed later.

A useful comparison for the situation in the CHTSC is found in superfluid liquid He with a condensation energy of 2 K ~ 0.2 meV. In this case, the coupling energy of the "boson", as measured by the $^1S_0 \rightarrow {}^2S_{1/2}$ excitation of the ground state of the He atom, is 25 eV, much larger than the condensation energy. The situation is sketched in Fig. 1 with reference to the CHTSC. In Fig. 1a, the doping dependence of the gaps is shown for a CHTSC with $T_c$ = 95 K (such as Y 123, Bi 2212 or Hg1201 at OP). For x = 0.27, the ratio of the pseudogap and the superconducting gap energy approaches 1, representing the classical BCS case. This is a special situation because in this case the coupling energy is equal to the condensation energy. On the other side of the phase diagram, i.e., at the minimum doping at x = 0.05, the ratio of the coupling energy (straight line) and the gap energy (parabola) approaches infinity, representing the rigorous BEC case.

In Figs. 1b) and 1c), spectroscopic results for the excitation energies are shown schematically. In the BCS case, this is the gap measured by tunnelling spectroscopy or ARPES experiments (as, e.g., in Nb metal). Unfortunately, such a simple example cannot be given for the BEC case, because (to our

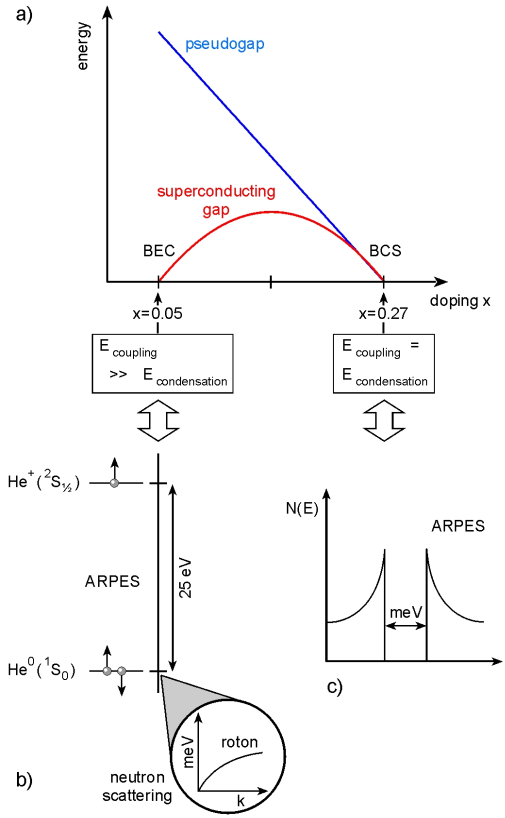

**Figure 1** (color online):
Schematic drawing of the phase diagram of a CHTSC. a) Gap-vs.-hole doping (x) with linear doping dependence of the pseudogap and parabolic doping dependence of the superconducting gap. At x = 0.05, there is the canonical BEC case with $E_{coup}/E_{cond} \to \infty$, at x = 0.27, there is the canonical BCS case with $E_{coup}/E_{cond} = 1$; b) illustration of the BEC situation for liquid He. The boson binding energy corresponds to the $^1S_0 \to {}^2S_{1/2}$ excitation energy (~ 25 eV). The condensate excitation is in the meV regime as measured by neutron scattering [14]; c) illustration of the excitation of a BCS system (e.g., Nb at 1 K) where the gap, as measured by ARPES or tunnelling, reflects the boson coupling energy and the condensation energy [14].

knowledge) there exists no canonical electronic BEC system up to now. Well documented BEC systems are liquid He and the ultra cold atomic gases. In Fig. 1b), the situation is sketched for liquid He. Photoelectron spectroscopy on He can measure the "break-up energy" of the $(1s^2)$ boson (in the present context, this is the pseudogap) which is 25 eV, leaving a He ion in a $^2S_{1/2}$ final state (with one electron instead of two electrons). The condensation energy has been measured by INS [14]. It is in the meV regime and corresponds to the superconducting gap in the present context.

Looking at Fig. 1a), one realizes that for the CHTSC, there are a pseudogap and a superconducting gap present for all dopings. In this picture, the pseudogap is the coupling gap and the superconducting gap is the condensation gap. In the BCS situation, the pairing gap is equal to the condensation gap and a one-electron spectroscopy measures these identical gaps. However, in the BCS-BEC cross-over situation, the pairing gap energy is larger than the condensation energy, meaning that $\Delta_{pg}$ is larger than $\Delta_{sc}$, as evident from Fig. 1a). This also means that in any situation between strict BCS and strict BEC there are always a pairing gap and in addition a superconducting gap.

The meaning of the superconducting gap has however to be qualified in the BCS-BEC cross-over situation. If in this case a one-electron spectroscopic measurement is performed, it measures the coupling gap (pseudogap) by breaking up the pair. The superconducting gap (condensation gap) is a two-electron excitation (exciting a boson out of the condensate) and can only be measured by two-electron spectroscopies, meaning that such experiments measure the total condensation energy $E_{sc}$ which can be written as $E_{sc} = 2\Delta_{sc}$. Therefore, if one talks about a superconducting gap in the BCS-BEC cross-over situation, one means half the condensation energy. It is the order parameter, that follows in its doping dependence that of $T_c$ and that disappears at $T_c$.

In the following discussion, the simple assumption is made, that the 70 meV signal as well as the 40 meV signal always represents the same excitation, irrespective of the experimental technique used to obtain them. This is by no means generally accepted. For example, the gap at $2\Delta = 70 \pm 10$ meV, as measured by ARPES, is assumed to be the pseudogap above $T_c$ and the superconducting gap below $T_c$ [35], and the same interpretation is given for this excitation as measured by STM [31]. In an other example, the $2\Delta = 40 \pm 5$ meV excitation is taken as the superconducting gap in ASJ experiments [3, 4] and a spin-spin correlation function in the INS experiment [45-49].

In this communication it will be shown, that the 40 meV excitation is the order parameter. Therefore, a method that is well established in measuring the superconducting gap, namely ASJ will be discussed and by its nature it can be compared in a phenomenological way to $B_{2g}$ RS and INS experiments leading to arguments that these techniques indeed measure the same 40 meV excitation, namely the order parameter.

### 3 Andreev Saint-James (ASJ) reflection

Although ASJ represents a tunnelling experiment, it is distinctly different from NIS or SIS tunnelling [30, 50]. This is already apparent in the measured signal. While NIS and SIS experiments show "no" signal in the range of $\pm\Delta_{sc}$ or $\pm 2\Delta_{sc}$, respectively, and a (normalized) conductance of 1 outside this energy values (an intuitively appealing result), ASJ shows a (normalized) conductance of 2 within the gap region (±$\Delta_{sc}$) and a (normalized) conductance of 1 outside this region. Thus, one can assume in a simplified picture, that in NIS and SIS the electron impinging (or being emitted from) the NS interface is unable to penetrate the energy range ±$\Delta_{sc}$ (or ±$2\Delta_{sc}$) of the superconductor (this is of course the meaning when talking of a gap in the superconductor). In contrast, in ASJ, the impinging (or emitted) electron is transformed into a boson at the interface (see Ref. 50 for a transparent discussion of this point) and now the gap region is transparent for it. Since a boson is formed out of an electron, the conductance therefore rises from 1 to 2.

According to Tinkham [30, 50] one can call NIS/SIS a one-electron process because one electron transverses the NS interface while ASJ can be called a two-electron process because two electrons transverse the NS interface. Of course, one can call all three tunnelling experiments one-electron processes because one electron is impinging on the interface. However we have decided to use Tinkham´s notation.

While this distinction between NIS/SIS and ASJ is trivial and in BCS superconductors all these techniques measure the same $\Delta_{sc}$, it is useful to have a further look at the situation for the CHTSC.

BCS is a special case since the pair breaking energy is equal to the condensation energy. From the above discussion it is obvious that NIS/SIS measure the pair breaking energy while ASJ measures the condensation energy. Therefore, if NIS/SIS and ASJ are applied to systems where the coupling energy is different from the condensation energy, as in the case of CHTSC [3, 4, 32-34], NIS/SIS and ASJ are expected to provide different results for the experimental gap, as they do in the CHTSC. In our opinion, this can be considered as the most convincing proof that the CHTSC do not represent BCS systems. In addition, one also realizes that the NIS/SIS gap is larger than the ASJ gap in agreement with the expectation for a BCS-BEC cross-over system (see Fig. 1).

It is worthwhile to look at ASJ reflection in more detail. In a normal-superconducting junction with zero interface strength Z [30, 50], an electron that impinges on the junction from the normal-metal side with an energy below the gap energy $\Delta_{sc}$, can, in principle, not enter the superconductor, because there are no allowed states available for this electron. In other words: there is no ordinary tunnelling. However, if this electron forms a boson by creating simultaneously a hole, with the same energy as the impinging electron and being reflected from the interface, this boson can propagate in the condensate, carrying a charge of 2e, an effect called ASJ reflection.

Now, ASJ in the CHTSC shows a gap that is distinctly smaller than the gap measured in a NIS or SIS tunnelling experiment [3, 4] on the same sample under similar conditions; additionally, in contrast to the NIS/SIS tunnelling gap [31], the ASJ gap follows the doping dependence of $T_c$. By the nature of the ASJ process, the impinging electron can only be transformed into a superconducting boson, as which it propagates in the superconductor, as long as its energy is smaller than the superconducting gap energy (more precisely: smaller than half of the superconducting energy

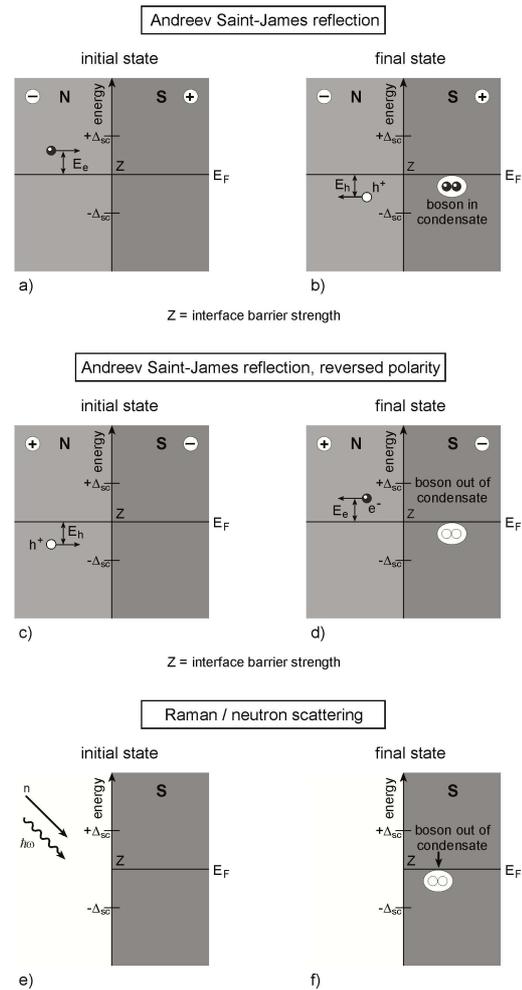

**Figure 2** (color online):
Principle of ASJ reflection, RS ($B_{2g}$) and INS in a CHTSC. a) initial state of ASJ with an electron impinging in the N-material on the N-S interface, the (+) and (-) sign indicate the polarity of the voltage applied to the sample; b) final state of ASJ with reflected hole in N-material and boson in the condensate in the superconducting state; c) same as a) with reversed polarity; d) same as b) with reversed polarity; e) initial state of $B_{2g}$ RS/INS with photon/neutron impinging on a S-material; f) final state of $B_{2g}$ RS/INS with a pair hole in the condensate, and a pg pair in the non-superconducting state of the electronic liquid. Note, that for instructive purpose also for the RS/INS experiment an N-S sample is shown with the N side being of course redundant in this experiment.

$E_{sc}$). If the energy of the electron is larger than the gap energy this impinging electron can propagate as an electron in the superconductor or transform into a pg pair which, however, dos not show a supercurrent [50].

One has to make an additional remark. Because, in the CHTSC one has to distinguish three regions with respect to the electronic structure: for $T>T^*$, one has a Fermi liquid, for $T_c>T>T^*$ one has a liquid consisting of electrons and pseudogap pairs, and for $T<T_c$ the electronic liquid consists of superconducting bosons, pseudogap bosons and electrons. In the second regime, there is a gap present (as seen by STM or NIS tunnelling). However, the creation of a pseudogap pair in a Andreev type experiment in region 2 will not lead to an enhancement of the current, because the pseudogap bosons do not create a supercurrent.

ASJ experiments, being two-electron experiments in the sense, that in the experiment two electrons transverse the barrier [30, 50], measure the width of the superconducting gap in the system under investigation, or the energy by which a boson is bound in the condensate. The comparison to STM and SIN experiments which are experiments evolving out of the ASJ case by increasing gradually the interface resistance Z (see Fig. 7 in ref.50) shows, that in the CHTSC this energy is different from the pseudogap energy. Therefore, the often expressed opinion that in the superconducting state of the CHTSC the superconducting gap evolves smoothly out of the pseudogap is hard to rationalize in view of an ASJ experiment. The two gaps are different properties in the superconducting state as suggested in many experiments [3, 4, 7]. In order to make now the connection of ASJ reflection to $B_{2g}$ RS and INS experiments (magnetic resonance), the principles of these experiments are sketched in Fig. 2. Quite artificially, the experiments have been broken-up into an initial and a final state. While Figs. 2a) and 2b) reflect the textbook description of ASJ [50], the actual experiment is symmetric around zero-potential (U=0). The case of the applied potential reversed relative to Figs. 2a) and 2b) is shown in Figs. 2c) and 2d). In principle, the velocities of all particles have to be reversed relative to the situation shown in Figs. 2a) and 2b). This means that a hole is impinging on the interface in the initial state, and in addition an electron is travelling away from the interface and a "boson hole" is created in the final state. This latter situation will be useful in dealing with $B_{2g}$ RS and INS.

There is another subtle point worth to be mentioned. ASJ reflection is monitored by the energy of the impinging electron or hole, and if $|E_e|=\Delta_{sc}$, seemingly the superconducting gap is measured. By the very nature of this technique, however, in addition the energy of the hole (electron) is measured and, in principle, ASJ reflection therefore does not measure $\Delta_{sc}$, but $2\Delta_{sc} = E_{sc}$.

The case for this statement can be made also differently [50]. The electrical power at the interface can be written as

$$P = \frac{U^2 \cdot A \cdot \sigma}{\ell}$$

(U: applied voltage, A: area, $\ell$: length of the sample, $\sigma$: conductance. Now, if one is working above $T_c$, one has for the normal state conductance

$$\sigma_{N-N} = R_N \cdot dI/dU = 1.$$

In the case, where the superconductor is below $T_c$ one has in the ASJ regime

$$\sigma_{N-S} = R_N \cdot dI/dU = 2 \ (U \leq \Delta_{sc}).$$

Applying a voltage of magnitude $\Delta_{sc}$ to the interface in the normal state leads to

$$P_{N-N} = \frac{A}{\ell} \cdot \Delta_{sc}^2 \cdot \sigma_{N-N} = \frac{A}{\ell} \cdot \Delta_{sc}^2,$$

while at the normal-superconducting interface the power is

$$P_{N-S} = \frac{A}{\ell} \cdot \Delta_{sc}^2 \cdot \sigma_{N-S} = \frac{A}{\ell} \cdot \Delta_{sc}^2 \cdot 2 = \frac{A}{\ell} \cdot \left(\Delta_{sc}^2 + \Delta_{sc}^2\right).$$

Therefore, ASJ measures twice the gap which is a complicated way of stating that for $U \leq \Delta_{sc}$ it has a conductance of twice the normal state conductance [50].

## 4 Raman scattering

Raman scattering on CHTSC [28] is commonly performed in two symmetries, namely $B_{1g}$ and $B_{2g}$, leading to two different excitations (there is in addition an $A_{1g}$ excitation for which the interpretation is however not yet established [28]). The excitation in the $B_{1g}$ symmetry (probing the ($\pi$/0) region of the Brillouin zone) has a linear doping dependence as the excitation measured by ARPES or STM. Its energy is twice the value of that obtained by, e.g., ARPES experiments.

In the Raman literature [25-28] it is argued using the comparison in energy and doping dependence to that found in ARPES and NIS/SIS experiments, that the $B_{1g}$ signal observed is due to a pseudogap excitation. Since the energy of the RS $B_{1g}$ signal is twice that of the ARPES signal one can conclude that the $B_{1g}$ RS signal measures a pairing gap.

The excitation in the $B_{2g}$ symmetry has a parabolic doping dependence, similar to that of $T_c$. Therefore, it can be considered the order parameter.

The question of course arises about the nature of the order parameter. This is established in BCS systems, where the order parameter is the pair breaking or condensation energy. Guyard et al. [26] follow the BCS picture in calling the $B_{2g}$ RS signal a measurement of the pairing gap and correspondingly give a comparison with BCS theory. However, the situation is more difficult in a situation like that in Fig. 1b) which, although with very different numbers, applies to the CHTSC. Here, the pairing gap is obviously not the order parameter, and the order parameter, as measured by, e.g., INS, is not a pairing gap. With respect to this diagram, one can argue [14] that the order parameter is the energy needed to take a boson (in the case of Fig. 1b) a He atom) out of the superfluid condensate. Now, one can assume that a similar interpretation is also valid for the CHTSC. This means that the $B_{1g}$ signal measures the pairing energy and the $B_{2g}$ signal measures the order parameter, namely the energy needed to take a boson out of the condensate [26]. Its magnitude is identical to that measured by ASJ reflection [3, 4]. The $B_{2g}$ Raman excitation can be considered as a reverse ASJ experiment, namely as the excitation of a superconducting boson out of the condensate, see Figs. 2e) and 2f).

Standard Raman theory used to explain the data on the CHTSC [25, 28], gives a different picture because it assumes a BCS type state for the CHTSC. A particular transparent discussion of Raman scattering in the CHTSC along this line is found in the work of LeTacon et al. [25] They show that the current BCS based theory of Raman scattering in CHTSC relies on the assumption of one gap and "fails to reproduce the opposite doping dependence of the antinodal and the nodal peaks". These authors come to the conclusion that probably two different energy scales are needed to explain the Raman data [26, 27]. This is exactly what is suggested here in identifying the $B_{1g}$ resonance with the pseudogap excitation and the $B_{2g}$ resonance with the order parameter.

## 5 Neutron scattering

In principle, neutron scattering is often the most useful technique for the investigation of solid state properties, and therefore, it has also been applied intensively to the CHTSC [29, 45-49]. Probably the most important information that has come out of these investigations is the so-called spin resonance with an "hourglass" dispersion, providing an energy of ~ 40 meV (for systems with $T_c$ = 95 K) at (0.5, 0.5) rlu. As a function of temperature this resonance gets smaller if $T_c$ is approached from below. Thus, from this point of view, it could be regarded as the order parameter. Another indication that the order parameter is measured is given by the doping dependence of the resonance energy because it follows that of $T_c$.

However, the standard interpretation of the resonance energy is a different one, namely that of a spin correlation function.

An interesting point for INS data of CHTSC is the following. The parent compounds of the CHTSC are antiferromagnets. For appropriate boundary conditions, a piece of an antiferromagnet exhibits no magnetic moment. Doping the antiferromagnet can be regarded as replacing an oxygen atom with $p^6$ configuration by an oxygen ion with $p^5$ configuration. The $p^5$ configuration has a net spin ½ which couples antiferromagnetically to the spin of the adjacent $Cu^{2+}$ ion, forming a Zhang Rice singlet (ZRS). This ZRS has annihilated one spin of the antiferromagnet (although it sounds strange with respect to the term "singlet", a ZRS creates a net spin in the antiferromagnetic host lattice). This net spin can couple to the spin of a neutron, giving rise to the signal in neutron scattering. Assuming that in the superconducting state the charge carriers are condensed ZRSs the arguments given above show, that neutrons can couple to the bosons in the condensate, and thus INS is able excite these bosons.

A particular feature of the neutron data is the dispersion of the resonance at (0.5/0.5) which has been called x-like or hourglass-like. The reason for this designation can be seen from Fig. 3. This figure contains dispersion measurements on Y 123 from four different sources, where three [45-47] used single crystals with quite similar doping and correspondingly similar $T_c$ (namely 59 K, 61 K and 63 K). A fourth experiment [48] was performed on a sample with $T_c$ = 89 K. In order to compare the different data within one diagram the energies are given in fractions of the resonance energy at (0.5/0.5). There is a considerable scatter in the data, also between those data sets from the three samples with the transition temperatures around 60 K.

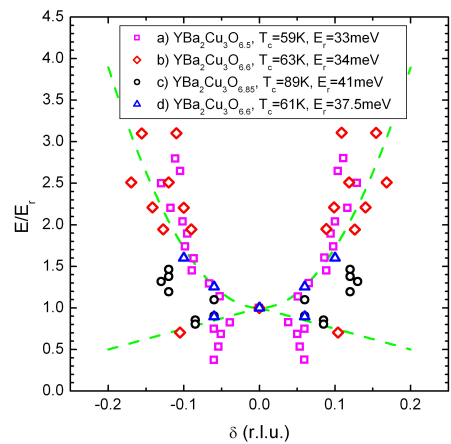

**Figure 3** (Color online)
Neutron scattering results for Y-Ba-Cu samples with different dopings: (a) Ref. 47, (b) Ref. 46, (c) Ref. 48, (d) Ref. 45. On the ordinate, the energies are given in fractions of the resonance energy at (0.5/0.5). Theoretical curves are a parabola for E > E(0.5/0.5) and a straight line for E > E(0.5/0.5).

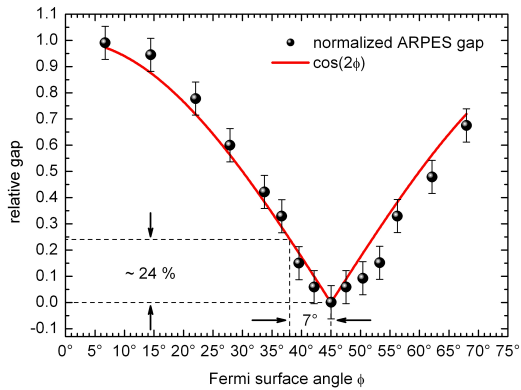

**Figure 4** (Color online)
Gap dispersion for a $T_c$ = 87 K Bi 2212 sample; a deviation of 7° from the node leads to a reduction in the gap of 24%. The gap energies are given in relation to the value at (π/0) [35, 51].

The data above $E_r$ can be fitted by a parabola with its bottom at $E_r$, as shown in Fig. 3. This is in agreement with the present interpretation of the neutron data, because above $E_F$ the bosons represent "free particles", having a parabolic dispersion as a function of the wave number.

The dispersion below $E_r$ can be interpreted as follows: A change of the wave vector by Δ ~ 0.1 in rlu. would result in a change of 7° in the direction of the wave vector excited by the neutron. In terms of the d-gap shown in Fig. 4 [51] this causes a reduction of the energy of the boson by about 25%, where the boson energy dispersion is close to linear in this region of the cosine dependence. This would also result in a linear dependence of the neutron resonance energy (which we here identify with the superconducting boson) dispersion below $E_r$, as indicated by the straight line in Fig. 3. While this seems to be in agreement with the data from three investigations, it disagrees with the results of Stock et al. [47]. So, at this point, there obviously remains a problem: it can be in the proposed interpretation but also with the deviation of the data sets of different groups.

By the arguments just presented, INS should also see the pseudogap excitations, because in the picture used here the pseudogap bosons are the preformed pairs. There is however one difference between the bosons in the superconducting state and those in the pseudogap state. The latter are relatively short lived, as can be seen, e.g., from the large ARPES linewidths [35], while the bosons in the condensate are long lived. The experiments on the magnetic resonance peak at 40 meV show only a relatively weak signal. Therefore it can be possible that the observation of the pseudogap by INS has so far escaped from observation due to intensity reasons.

However a recent INS experiment on $La_{2-x}Sr_xCuO_4$ [52] has probably seen the pseudogap excitation, supporting our conjecture. However this type of experiment has to be extended to other systems, in order to verify its general validity

There are different interpretations of the neutron resonance and many of them have been recently reviewed by Eschrig [53] (see also Refs. [54-56]). Many of these interpretations view Er as a spin excitation to a spin 1, where however the theories differ in importat details. The calculations start from a generalized susceptibility [53-57]. $E_r$ is than obtained as a pole in the susceptibility. The interpretations of $E_r$ obtained in this way can approximately be grouped into two classe. Most authors prefer a two particle analysis (particle-hole or particle-particle) while Onufrieva and Pfeuty [56] point out, that their results are very different from the two particle scenarios in that they find, that $E_r$ is dominated by collective effects and they state that the physics according to their results is similar to that of liquid helium, with its roton mode. There is a problem with these calculations. Because they start with a susceptibility in the BCS approximation, they use only one gap and this is the one obtained experimentally by, e.g., ARPES, which is linearly dependent on doping. There is much evidence that this gap is the pseudogap while the superconducting gap is different from it and measured e.g. by ASJ. If one would insert the ASJ gap into the calculations of ref. [53-56], they would all yield a resonance energy not in agreement with experiment. In essence the question, whether the current spin-1 exciton theories describe the INS measured $E_r$ correctly depends on the picture that is used for the interpretation of the measured gaps in the CHTSC. If one assumes that the pseudogap is the only gap and that it is identical to the superconducting gap below $T_c$, than the spin-1 exciton theories describe the neutron resonance correctly (apart from the different interpretation as to whether it is a two particle or a collective resonance). If however one assumes that the pseudogap state is a precursor state to the superconducting state and the superconducting gap is measured eg. by ASJ or $B_{2g}$ RS or INS, than the current theories to explain the INS resonance have to e looked upon with caution.

There is yet another approach to explain the magnetic excitations in the CHTSC which relies on fluctuating stripe domains ( and checkerboards) [58].

Thus it seems fair to state that there is no consistent theoretical approach for an analysis of the INS data in the CHTSC.

# 6 Discussion

The present analysis of ASJ $B_{2g}$ RS and INS data supports the emerging picture of the CHTSC [2, 3, 7, 15-22]. In this concept, the superconducting state is derived from the doping produced singlets of the original antiferromagnetic insulator, cf. Q(i). The energy of these singlet states, measured by the pseudogap energy, decreases with increasing doping due to an increasing screening. The superconducting transition temperature is given by the product of the charge carrier density and the boson coupling strength (pseudogap energy) [7, 22-24], cf. Q(iii). Since the two terms of this product have an approximately linear doping dependence, the order parameter (or superconducting transition temperature) has a parabolic dependence on doping, as observed in experiment [24]. This picture shows that the two gaps are a necessary ingredient of the CHTSC, and that the pseudogap is not the signature of a competing phase, cf. Q(iv). This concept suggests that the CHTSC are in the BCS-BEC cross-over regime [5, 13, 15], cf. Q(ii).

It is interesting that this general picture was developed by Schmitt-Rink and Nozieres [15] before the discovery of the high temperature superconductors. This paper also shows that the BCS superconductors are a special case, where the coupling energy of the boson is in addition the order parameter. The more general case is that, where the coupling strength is larger than the order parameter which gives a wide cross-over regime from the BCS case to the case of the BEC [13, 14], where the coupling strength of the boson is very large compared to the order parameter.

In the photoemission and the tunneling community very often a one-gap situation is favoured. This is, e.g., expressed by Fischer et al. [31] who state *"the pseudogap and the superconducting gap have a comparable magnitude and scale with each other as a function of doping"*, and, *"as the temperature is raised from $T < T_c$ the superconducting gap remains constant and smoothly evolves into a pseudogap across $T_c$"*. These statements result from a BCS-like view for the CHTSC. However if the coupling energy is larger than the condensation energy, the one electron spectroscopies like STM or ARPES measure the coupling energy but cannot measure the smaller condensation energy (for a qualification of this statement, see Refs.3 and 7). This is actually supported convincingly by tunnelling experiments in high magnetic fields [31]. If the tip in these experiments is swept through a vortex *"the spectra evolve in the same way as when the temperature is raised above $T_c$"*. Again, this is in agreement with the suggestion that STM measures the pseudogap below and above $T_c$ : in the core there is a non-superconducting region and there the electronics state is similar to that in the state made non superconducting by raising the temperature (pseudogap state). This seems to be evidence for the coexistence of the pseudogap and the superconducting gap below $T_c$.

While the pseudogap is well documented in the literature (see, e.g., Damascelli et al. [35]) the situation with respect to the superconducting gap is less clear. Therefore, it was demonstrated here that ASJ, $B_{2g}$ RS and INS measure the superconducting gap which is distinctly different from the pseudogap, cf. Q(v). It is important to realize that while it is convenient to argue in terms of the superconducting gap $\Delta_{sc}$, the property actually measured is $E_{sc} = 2\Delta_{sc}$.

By referring to Fig. 1b), the pseudogap is a pairing gap and it is measured by breaking up the pair, as, e.g., in ARPES or STM, but also in $B_{1g}$ RS. However, in contrast to the BCS situation, this is not the order parameter in the CHTSCs. The order parameter is given by the energy that distinguishes the superconducting state from the non-superconducting state and is therefore the condensation, measured by taking a boson out of the condensate or putting the boson into it.

# 7 Conclusions

In this communication, arguments have been presented that ASJ, $B_{2g}$ RS and INS (magnetic resonance) measure the same property, namely the order parameter of the CHTSC. In addition, it is suggested that the "hourglass"-like dispersion seen in the INS experiments has a parabolic form for $E > E_r$ and a linear form for $E < E_r$.

On a broader scale it is also suggested that the eight experiments (ARPES, NIS, SIS, STM, RS $B_{1g}$, thermodynamics, NMR and nodal ARPES "kink") that measure a 70 meV excitation in a prototype $T_c$ = 95 K system (such as Bi2212, Y123 or Hg1201) all probe the same property: It is the $(\pi,0)$ pseudogap, having a linear doping dependence and representing the pairing energy of the bosons in the CHTSC. In addition, the seven experiments (ASJ, INS, optics, $B_{2g}$ RS, ARPES "dip", NIS/SIS "dip" and antinodal ARPES "kink") that measure a 40 meV excitation in the same prototype materials also probe the same property: the condensation energy of the superconducting condensate.

This paper has benefited very much from discussions with B. Keimer. S.H. thanks to the University of British Columbia (Vancouver) for its hospitality where this work was initiated. The discussions with A. Damascelli and G.A. Sawatzky have been very helpful.

**References**

1. J. G. Bednorz, K. A. Müller, Z. Phys. B 64, 189 (1986)
2. B. Eddeger, V.N. Muthukuman, C. Gros, Adv. Phys. **56** (2007) 927
3. G. Deutscher, Rev. Mod. Phys. **77**, 109 (2005)
4. G. Deutscher, Nature **397**, 410 (1999)
5. Y.J. Uemura, Physica B, **374**,1, (2006)
6. M.R. Norman, D. Pines, C. Kallin, Advances in Physics **54**, 715 (2005)
7. S. Hüfner, M. Hossain, A. Damascelli, G. Sawatzky, Rep. Prog. Phys 71 (2008) 062 501
8. S. Hüfner, M.A. Hossain, F. Müller, Phys. Rev. B 78, 014519 (2008)
9. S. Hüfner, F. Müller, Phys. Rev. B 78, 014521 (2008)
10. B. Goss Levi, Physics Today **60**, 17 (2007)
11. A. Cho, Science **314**, 1072 (2006)
12. A.J. Millis, Science **314**, 1888 (2006)
13. Q. Chen, J. Stajic, S. Tan, K. Levin, Physics Reports 412, 1 (2005)
14. A.J. Leggett, *Quantum Condensates,* Oxford University Press (2006)
15. P. Nozieres, S. Schmitt-Rink, J. Low Temp. Phys. 59, 195 (1985)
16. P.W. Anderson, Science 235, 1196 (1987)
17. A. Paramekanti, M. Randeria, N. Trivedi, Phys. Rev. B 70, 054504 (2004)
18. G.Kotliar, J.Liu, Phys. Rev. B 38, 5142 (1988)
19. C. Gross, Phys. Rev. 38, 931 (1988)
20. F.C. Zhang, T.M. Rice, Phys. Rev. B 37, 3759 (2003)
21. H. Eskes, L. H. Tjeng, and G. A. Sawatzky, Phys. Rev. B 41, 288 (1990)
22. V.J. Emery, S.A. Kivelson, Nature 374, 434 (1995)
23. C.M. Varma, Phys. Rev. B 73, 155113 (2006)
24. M. R. Presland, J. L. Tallon, R. G. Buckley, R. S. Liu, N. E. Flower, Physica C 176, 95 (1991)
25. M. Le Tacon, A. Sacuto, A. Georges, G. Kotliar, Y. Gallais, D. Colson, A. Forget, Nature Physics, **2**, 537 (2006)
26. W. Guyard, A. Sacuto, M. Cazayous, Y. Gallais, M. Le Tacon, D. Colson, and A. Forget, Phys. Rev. Lett. **101**, 097003 (2008)
27. W. Guyard, M. Le Tacon, M. Cazayous, A. Sacuto, A. Georges, D. Colson, and A. Forget, Phys. Rev. B **77**, 024524 (2008)
28. T.P. Devereaux, R. Hackl, Rev. Mod. Phys. 79, 175 (2007)
29. Y. Sidis, S. Pailhès, B. Keimer, P. Bourges, C. Ulrich, L. P. Regnault, Z. Phys. Stat. Sol. B 241, 1204 (2004)
30. M. Tinkham, *Introduction to Super- conductivity*, 2nd edition (Dover Publishers, Minneola, N.Y. 1996)
31. O. Fischer, M. Kugler, I. Maggio-Aprile, C. Berthod, C. Renner, Rev. Mod. Phys.79, 353 (2006)
32. N. Miyakawa, P. Guptasarma, J.F. Zasadzinski, D.G. Hinks, K.E. Gray, Phys. Rev. Lett. 80, 157 (1998)
33. N. Miyakawa, J.F. Zasadzinski, L. Dzuzu, G. Guptasarma, D.G. Hinks, C. Keudziora, K.G. Gray, Phys. Rev. Lett. 83,1018 (1999)
34. J.F. Zasadzinski, L. Ozyuzer, N. Miyakawa, K.E. Gray, D.G. Hinks, C. Kendzora Phys. Rev. Lett. 87, 067005 (2001)
35. A. Damascelli, Z. Hussain, Z.X. Shen, Rev. Mod. Phys. **75**, 473 (2003)
36. A.A, Kordyuk, S.V. Borisenko, V.B. Zabolotnyy, J. Geck, M. Knupfer, J. Fink, B. Büchner, C. T. Lin, B. Keimer, H. Berger, A.V. Pan, S. Komiya, Y. Ando, Phys. Rev. Lett. 97, 017002 (2006)
37. S.V. Borisenko, A.A. Kondyuk, V. Zaabolotnyy, J. Geck, D. Inosov, A. Koitsch, J. Fink, M. Knupfer, B. Büchner, V. Hinkov, C.T. Liu, B. Keimer, T. Wolf, S.G. Chiuzbrian, L. Patthey, R. Follack, Phys. Rev. Lett. 96, 117004 (2006)
38. T. Timusk, B.W. Statt, Rep. Prog. Phys. 62, 61 (1999)
39. A. Kanigel, U. Chatterjee, M. Randeira, M. R. Norman, G. Koren, K. Kadowaki, J.C.Campuzano, Phys. Rev. Lett. 101, 137002(2008)
40. D.N. Basov, T. Timusk, Rev. Mod Phys. 77, 721 (2005)
41. J. Yang,J. Hwang, E. Schachinger, J. B. Carbotte, R.P.S.M. Lobo, D. Colson, A.Forget, T. Timusk, Phys. Rev. Letters, 102, 0273003 (2009)
42. J. C. Campuzano, H. Ding, M.R. Norman, H. M. fretwell, M. Randeira, A. Kaminski, J. Mesot, T. Takeuchi, T. Sato, TYokoya, T.Takahashi, T. Mochiku, Phys. Rev. Letters.,83 ,3709, 1999)
43. J. Fink, A. Koitsch, J. geck,V. Zababalotnyy, M. Knupfer, B. Büchner, A. Chubukov, H. berger, Phys. Rev. B 74, 165102 (2006)
44. A.D. Gromko, A. V. Fedorov, Y. D. Chuand., J. D. Koralik, Y. Aiura, Y.Yamaguchi,

K.Oka,Y.Ando,D.S. Dessau, Phys. Rev B 68, 174520 (2003)
45. V. Hinkov, P. Bourges, S. Pailhès, Y. Sidis, A. Ivanov, C. Frost, T. Perring, C. Lin, D. Chen, B. Keimer, *Nature Physics 3, 780 (2007)*
46. S.M. Hayden, H.A. Mook, P. Dai, T.G. Perring, F. Dogan, Nature **429,** 531 (2004)
47. C. Stock, W. J. L. Buyers, R. A. Cowley, P. S. Clegg, R. Coldea, C. D. Frost, R. Liang, D. Peets, D. Bonn, W. N. Hardy, R. J. Birgeneau Phys. Rev. B **71**, 024522 (2005)
48. S. Pailhès, Y. Sidis, P. Bourges, V. Hinkov, A. Ivanov, C. Ulrich, L. P. Regnault, B. Keimer, Phys. Rev. Lett. **93**, 167001 (2004)
49. B. Keimer, private communication. We thank Prof. Keimer for extensive discussion on the neutron data
50. G. E. Blonder, M. Tinkham, and T. M. Klapwijk, Phys. Rev. B **25**, 4515 (1982)
51. H. Ding, M.R.Norman, J.C. Campuzano, M. Randeira, A.F. Bellmann, T.Yokoya, T. Takahashi, T.Mochiku, K. Kadowaki, Phys. Rev. B 54, R9678 (1996)
52. O. J. Lipscombe, B. Vignolle, T. G. Perring, C. D .Frost, S. M. Hayden, Phys Rev. Letters, 102 167002 (2009)
53. M. Eschrig, Advances in Physics 55,47 (2006)
54. P. A. Lee, N. Nagaosa, X-G. Wen, Rev. Mod. Phys.,78, 17 (2006)
55. J. Brinkmann and P. A. Lee Phys. Rev B 65,0114 502 (2001)
56. E. Onufrieva and P. Pfeuty, Phys Rev. B 65, 054 515 (2002)
57. J. R. Schrieffer, Theory of Superconductivity ,W.A. Benjamin, New York (1964)
58. M. Vojta, T. Vojta, R.K. Kaul, Phys. Rev. Letters, 97,097001 (2006)